# Corn Yield Prediction with Ensemble CNN-DNN

Mohsen Shahhosseini[1], Guiping Hu[1*], Saeed Khaki[1], Sotirios V. Archontoulis[2]

[1] Department of Industrial and Manufacturing Systems Engineering, Iowa State University, Ames, Iowa, USA
[2] Department of Agronomy, Iowa State University, Ames, Iowa, USA
* Corresponding author: E-mail: gphu@iastate.edu

## Abstract

We investigate the predictive performance of two novel CNN-DNN machine learning ensemble models in predicting county-level corn yields across the US Corn Belt (12 states). The developed data set is a combination of management, environment, and historical corn yields from 1980-2019. Two scenarios for ensemble creation are considered: homogenous and heterogenous ensembles. In homogenous ensembles, the base CNN-DNN models are all the same, but they are generated with a bagging procedure to ensure they exhibit a certain level of diversity. Heterogenous ensembles are created from different base CNN-DNN models which share the same architecture but have different levels of depth. Three types of ensemble creation methods were used to create several ensembles for either of the scenarios: Basic Ensemble Method (BEM), Generalized Ensemble Method (GEM), and stacked generalized ensembles. Results indicated that both designed ensemble types (heterogenous and homogenous) outperform the ensembles created from five individual ML models (linear regression, LASSO, random forest, XGBoost, and LightGBM). Furthermore, by introducing improvements over the heterogenous ensembles, the homogenous ensembles provide the most accurate yield predictions across US Corn Belt states. This model could make 2019 yield predictions with a root mean square error of 866 kg/ha, equivalent to 8.5% relative root mean square and could successfully explain about 77% of the spatio-temporal variation in the corn grain yields.
The significant predictive power of this model can be leveraged for designing a reliable tool for corn yield prediction which will in turn assist agronomic decision makers.

# 1. Introduction

Accurate crop yield prediction is essential for agriculture production, as it can provide insightful information to farmers, agronomists, and other decision makers. However, this is not an easy task, as there is a myriad of variables that affect the crop yields, from genotypes, environment, and management decisions to technological advancements. The tools that are used to predict crop yields are mainly divided into simulation crop modeling and machine learning (ML).

Although these models are usually utilized separately, there have been some recent studies to combine them towards improving prediction. The outputs of crop models have served as inputs to multiple linear regression models in an attempt to make better crop yield predictions (Mavromatis, ,2016; Busetto et al., 2017; Pagani et al., 2017). Some other studies have made additional advancement and created hybrid crop model-ML methodologies by using crop model outputs as inputs to a ML model (Everingham et al., 2016; Feng et al., 2019). In a recent study, Shahhosseini et al., (2021) designed a hybrid crop model-ML ensemble framework, in which APSIM was used to provide additional inputs to the yield prediction task. The results demonstrated that coupling APSIM and ML could improve ML performance up to 29% compared to ML alone.

On the other hand, the use of more complex machine learning models with the intention of better using numerous ecological variables to predict yields has been recently becoming more prevalent (Basso and Liu, 2019). Although there is always a tradeoff between the model complexity and its interpretability, the recent complex models could better capture all kinds of associations such as linear and nonlinear relationships between the variables associated with the crop yields, resulting in more accurate predictions and subsequently better helping decision makers (Chlingaryan et al., 2018). These models span from models as simple as linear regression, k-nearest neighbor, and regression trees (González Sánchez et al., 2014; Mupangwa et al. ,2020), to more complex methods such as support vector machines (Stas et al., 2016), homogenous ensemble models (Vincenzi et al., 2011; Fukuda et al., 2013; Heremans et al., 2015; Jeong et al., 2016; Shahhosseini et al., 2019), heterogenous ensemble models (Cai et al., 2017; Shahhosseini et al., 2020; Shahhosseini et al., 2021), and deep neural networks (Liu et al., 2001; Drummond et al., 2003; Jiang et al., 2004; Pantazi et al., 2016; You et al., 2017; Crane-Droesch, 2018; Wang et al., 2018; Khaki and Wang, 2019; Kim et al., 2019; Yang et al., 2019; Jiang et al., 2020; Khaki et al., 2020a; Khaki et al., 2020b). Homogeneous ensemble models are the models created using same-type base learners, while the base learners in the heterogenous ensemble models are different.

Although deep neural networks demonstrate better predictive performance compared to single layer networks, they are computationally more expensive, more likely to overfit, and may suffer from vanishing gradient problem. However, some studies have proposed solutions to address these problems and possibly boost deep neural network's performance (Bengio et al., 1994; Srivastava et al., 2014; Ioffe and Szegedy, 2015; Szegedy et al., 2015; Goodfellow et al., 2016; He et al., 2016).

Convolutional neural networks (CNNs) have mainly been developed to work with two-dimensional image data. However, they are also widely used with one-dimensional and three-dimensional data. Essentially, CNNs apply a filter to the input data which results in summarizing different features of the input data into a feature map. In other words, CNN paired with pooling operation can extract high-level features from the input data that includes the necessary information and has lower dimension. This means CNNs are easier to train and have fewer parameters compared to fully connected networks (Goodfellow et al., 2016; Zhu et al., 2018; Feng et al., 2020).

Since CNNs are able to preserve the spatial and temporal structure of the data, they have recently been used in ecological problems, such as yield prediction. Khaki et al. (2020b) proposed a hybrid CNN-RNN framework for crop yield prediction. Their framework consists of two one-dimensional CNNs for capturing linear and nonlinear effects of weather and soil data followed by a fully connected network to combine high-level weather and soil features, and a recursive neural network (RNN) that could capture time dependencies in the input data. The results showed that the model could achieve decent relative root mean square error of 9% and 8% when predicting corn and soybean yields, respectively. You et al. (2017) developed CNN and LSTM models for soybean yield prediction using remote sensor images data. The developed models could predict county-level soybean yields in the U.S. better than the competing approaches including ridge regression, decision trees, and deep neural network (DNN). Moreover, Yang et al. (2019) used low-altitude remotely sensed imagery to develop a CNN model. The experimental results revealed that the designed CNN outperformed the traditional vegetation index-based regression model for rice grain yield estimation, significantly.

Another set of developed models to capture complex relationships in the input raw data are ensemble models. It has been proved that combining well-diverse base machine learning estimators of any types, can result in a better-performing model which is called an ensemble model (Zhang and Ma, 2012). Due to their predictive ability, ensemble models have also been used recently by ecologists. Several heterogenous ensemble models including optimized weighted ensemble, average ensemble, and stacked generalized ensembles were created using five base learners, namely LASSO regression, linear regression, random forest, XGBoost, and LightGBM. The computational results showed that the ensemble models outperformed the base models in predicting corn yields. Cai et al. (2017) combined several ML estimators to form a stacked generalized ensemble. The back-testing numerical results demonstrate that their model's performance is comparable to the USDA forecasts.

Although these models have provided significant advances towards making better yield predictions, there is still a need to increase the predictive capacity of the existing models. This can be done by improving the data collections, and by the means of developing more advanced and forward-thinking models. The ensemble models are excellent tools that have the potential to turn very good models to outstanding predictor models.

Motivated by the high predictive performance of CNNs and ensemble models in ecology (Cai et al., 2017; You et al., 2017; Yung et al., 2019; Shahhosseini et al., 2020; Khaki et al., 2020b;

Shahhosseini et al., 2021), we propose a set of ensemble models created from multiple hybrid CNN-DNN base learners for predicting county-level corn yields across US Corn Belt states. Building upon successful studies in the literature (Shahhosseini et al., 2020; Khaki et al., 2020b), we designed a base architecture consisting of two one-dimensional CNNs and one fully connected network (FC) as the first layer networks, and another fully connected network that combined the outputs of the first-layer networks and made final predictions, as the second-layer network. Afterwards, two scenarios are considered for base learner generation: heterogenous and homogenous ensemble creation. In the heterogenous scenario, the base learners are neural networks with the same described architecture, but with different depth levels. On the contrary, the homogenous ensembles are created with bagging the same architecture and forming diverse base learners. In each scenario, the generated base learners are combined by several methods including simple averaging, optimized weighted averaging, and stacked generalization.

## 2. Materials and Methods

The designed ensemble framework uses a combination of historical yield and management data obtained from USDA NASS, historical weather and soil data as the data inputs. The details of the created data set and the developed model will be explained below.

### 2.1. Data Preparation

- *Data sources*

The main variables that affect corn yields are environment, genotype, and management. To this end, we created a data set that includes weather, soil, and management data considering 12 US Corn Belt states (Illinois, Indiana, Iowa, Kansas, Michigan, Minnesota, Missouri, Nebraska, North Dakota, Ohio, South Dakota, and Wisconsin). It is also noteworthy that since only some of the locations across US Corn Belt states are irrigated, to keep the consistency across the entire developed data set, we assumed that all farms are rainfed and didn't consider irrigation as a feature. The variables weekly planting progress per state and corn yields per county were downloaded from USDA National Agricultural Statistics Service (NASS, 2019). The weather was obtained from a reanalysis weather database based off of NASA Power (https://power.larc.nasa.gov) and Iowa Environmental Mesonet (https://mesonet.agron.iastate.edu). Finally, the soil data was created from SSURGO, a soil database based off of soil survey information collected by the National Cooperative Soil Survey (Soil Survey Staff, 2019). These variables are described below. Across 12 states, on average the data from 950 counties in total were used per year.

- *Planting progress (planting date):* 52 features explaining weekly cumulative percentage of corn planted within each state (NASS, 2019)
- *Weather*: Four weather features accumulated weekly (208 features), obtained from NASA Power and Iowa Environmental Mesonet.

- o   Daily minimum air temperature in degrees Celsius
- o   Daily maximum air temperature in degrees Celsius
- o   Daily total precipitation in millimeters per day
- o   Shortwave radiation in watts per square meter
- o   Growing degree days
- *Soil*: The soil features wet soil bulk density, dry bulk density, clay percentage, plant available water content, lower limit of plant available water content, hydraulic conductivity, organic matter percentage, pH, sand percentage, and saturated volumetric water content. All variables determined at 10 soil profile depths (cm): 0–5, 5–10, 10–15, 15–30, 30–45, 45–60, 60–80, 80–100, 100–120, and 120-150. (Soil Survey Staff, 2019)
- *Corn Yield*: Yearly corn yield in bushel per acre, collected from USDA-NASS (NASS, 2019).

- *Data pre-processing*

The following pre-processing tasks were performed on the created data set to make it prepared for training the designed ensemble models.

- Imputing missing planting progress data for the state North Dakota before the year 2000 by considering average progress values of two closest states (South Dakota and Minnesota).

- Removing out-of-season planting progress data before planting and after harvesting.

- Removing out-of-season weather features before planting and after harvesting.

- Aggregating weather features to construct quarterly and annually weather features. The features solar radiation and precipitation were aggregated by summation, while other weather features (minimum and maximum temperature) were aggregated by a row-wise average.

- The observations with the yield less than 10 bu/acre were considered as outliers and dropped from the data set.

- Investigating the historical corn yields over the time reveals an increasing trend in the yield values. This could be explained as the effect of technological advances, like genetic gains, management progress, advanced equipment, and other technological advances. Hence, a new input feature was constructed using the observed trends that enabled the models to account for the increasing yield trend.
    - o   *yield_trend*: this feature explained the observed trend in corn yields. A linear regression model using the training data was built for each location as the trends for each site tend to be different. The year ($YEAR$) and yield ($Y$) features served as the predictor and response variables of this linear regression model, respectively.

Then the predicted value for each data point ($\hat{Y}$) is added as a new input variable that explains the increasing annual trend in the target variable. The corresponding value for the observations in the test data set was estimated by plugging in their corresponding year in the trained linear regression models ($\hat{Y}_{i,test} = b_{0_i} + b_{1_i} YEAR_{i,test}$). The following equation shows the trend value ($\hat{Y}_i$) calculated for each location ($i$), that is added to the data set as a new feature.

$$\hat{Y}_i = b_{0_i} + b_{1_i} YEAR_i \tag{1}$$

- All independent variables were scaled to be ranged between 0 and 1.

## 2.2. Base Models Generation

We propose the following CNN-DNN architecture as the foundation for generating multiple base learners that serve as the inputs to the ensemble creation models. The architecture consists of two layers of deep neural networks.

*First layer:*

Due to the ability of CNNs in capturing the spatial and temporal dependencies that exist in the soil and weather data, respectively, we decided to build two separate set of one-dimensional CNNs for each of the weather (W-CNN) and soil (S-CNN) groups of features. Such networks have been used before in different studies and have been proved to be effective in capturing linear and nonlinear effects in the soil and weather (Ince et al., 2016; Borovykh et al., 2017; Kiranyaz et al., 2019). In addition, a fully connected network (FC1) was built that took planting progress, and other constructed features as inputs and the output is concatenated with the outputs of the CNN components to serve as inputs of the second layer of the networks.

Specifically, the first layer includes three network types:

1) <u>Weather CNN models (W-CNN)</u>:
   CNN is able to capture the temporal effect of weather data measured over time. In the case of the developed data set, we will use a set of one-dimensional CNNs inside the W-CNN component.
2) <u>Soil CNN models (S-CNN)</u>:
   CNN can also capture the spatial effect of soil data which is measured over time and on different depths. Considering the data set, we will use a set of one-dimensional CNNs to build this component of the network.
3) <u>Other variables FC model (FC1)</u>:
   This fully connected network can capture the linear and nonlinear effect of other input features.

*Second layer (FC2):*

In the second layer we used a fully connected network (FC2) that aggregates all extracted features of the first layer networks (W-CNN, S-CNN, and FC1), and makes the final yield prediction.

The architecture of the proposed base network is depicted in Figure 1. As it is shown in the figure, the W-CNN and S-CNN components of the network each are comprised of a set of CNNs that are in charge of one data input type and their outputs are aggregated with a fully connected network. For the case of W-CNN component, there are 5 CNNs for each weather data type (precipitation, maximum temperature, minimum temperature, solar radiation, and growing degree days). Similarly, 10 internal CNNs are designed inside S-CNN component for each of the 10 soil data types. The reason we decided to design one CNN for each data type is the differences in the natures of different data types and our experiments showed that separate CNNs for each data type could extract more useful information and will result in better final predictions. The two inner fully connected networks (FC_W and FC_S) both have one hidden layer with 60 and 40 neurons, respectively.

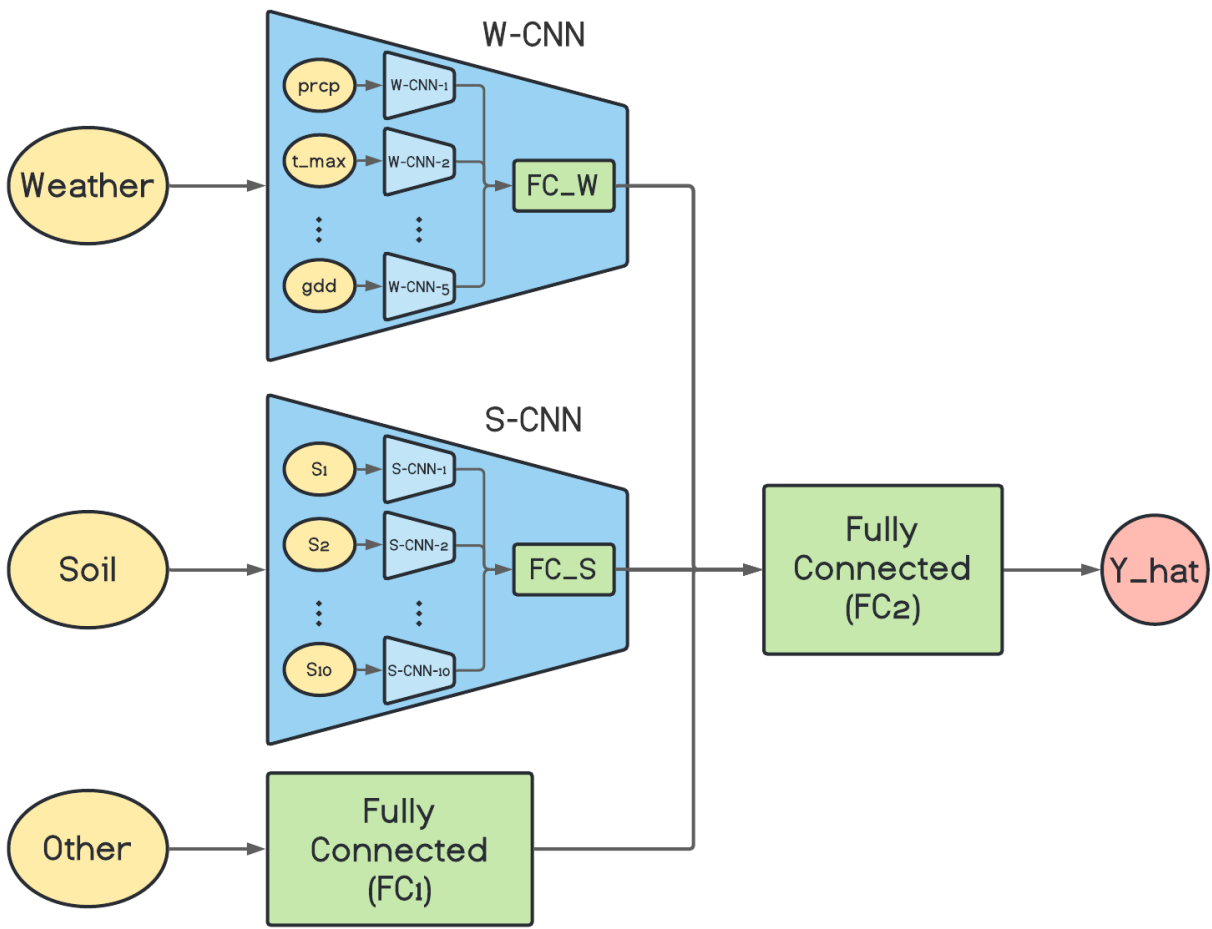

*Figure 1: The architecture of the proposed base network. prcp, t_max, and gdd represent precipitation, maximum temperature and growing degree days, respectively. S1, S2, … , and S10 are 10 soil variables which each are measured at 10 depth levels. Y_hat represents the final corn yield prediction made by the model.*

We used VGG-like architecture for the CNN models (Simonyan and Zisserman, 2014). The details about each of the designed CNN networks are presented in Table 1. We performed downsampling in the CNN models by average pooling with a stride of size 2. The feed-forward fully connected network in the first layer (FC1) has three hidden layers with 64, 32, and 16 neurons. The final fully connected network of the second layer (FC2) is grown with two hidden layers with 128 and 64 neurons. In addition, two dropout layers with dropout ratio of 0.5 are located at the two last layers of the FC2 to prevent the model from overfitting. We used Adam optimizer with the learning rate of 0.0001 for the entire model training stage and trained the model for 1000 iterations considering batches of size 16. Rectified linear unit (ReLU) was used as the activation function of all networks throughout the architecture except the output layer that had a linear activation function.

*Table 1: Detailed structure of the CNN networks of CNN components designed as the foundation for ensemble neural networks. The table on the left shows the details of the CNNs designed for each weather feature, and the right table presents the ones for the CNNs designed for each soil feature. FS, NF, S, and P represent filter size, number of features, stride, and padding.*

| CNNs in the W-CNN component | | | | |
|---|---|---|---|---|
| INPUT SIZE | $32 \times 1$ | | | |
| LAYER NAME | FS | NF | S | P |
| CONV1 | 6 | 4 | 1 | valid |
| AVERAGE POOLING 1 | 2 | - | 2 | valid |
| CONV2 | 3 | 4 | 1 | valid |
| AVERAGE POOLING 2 | 2 | - | 2 | valid |
| CONV3 | 3 | 4 | 1 | valid |
| AVERAGE POOLING 3 | 2 | - | 2 | valid |
| OUTPUT SIZE | $4 \times 1$ | | | |

| CNNs in the S-CNN component | | | | |
|---|---|---|---|---|
| INPUT SIZE | $10 \times 1$ | | | |
| LAYER NAME | FS | NF | S | P |
| CONV1 | 3 | 4 | 1 | valid |
| AVERAGE POOLING 1 | 2 | - | 2 | valid |
| CONV2 | 3 | 4 | 1 | valid |
| AVERAGE POOLING 2 | 2 | - | 2 | valid |
| CONV3 | 3 | 4 | 1 | valid |
| OUTPUT SIZE | $4 \times 1$ | | | |

To ensure that the ensemble created from a set of base learners performs better than them, the base learners should have a certain level of diversity and prediction accuracy (Brown, 2017). Hence, two scenarios for generating diverse base models are considered which are systematically different: homogenous and heterogenous ensemble base model generation.

- *Homogenous ensembles*

The homogenous ensembles are the models whose base learners are all the same type. Random forest and gradient boosting are examples of homogenous ensemble models. Their base learners are decision trees with the same hyperparameter values. Bootstrap aggregating (Bagging) is an ensemble framework which was proposed by Breiman (1996). Bagging generates multiple training data sets from the original data set by sampling with replacement (bootstrapping). Then, one base model is trained on each of the generated training data sets and the final prediction is the average (for regression problems) or voting (for classification problems) of the predictions made by each of those base models. Basically, by sampling with replacement and generating multiple data sets, and subsequently multiple base models, bagging ensures the base models have a certain level of diversity. In other words, bagging tries to reduce the prediction variance by averaging the predictions of multiple diverse base models.

Here, inspired by the way bagging introduces diversity in the base model generation, we design a bagging schema which generates multiple base CNN-DNN models using the same foundation model (Figure 1). This is shown in Figure 2. Then several ensemble creation methods make use of these bagged networks as the base models to create a better-performing ensemble network. We believe one drawback of bagging is assigning equal weights to the bagged models. To address that, we will use different ensemble creation methods in order to optimally combine the bagged models. We will discuss ensemble creation in the next chapter.

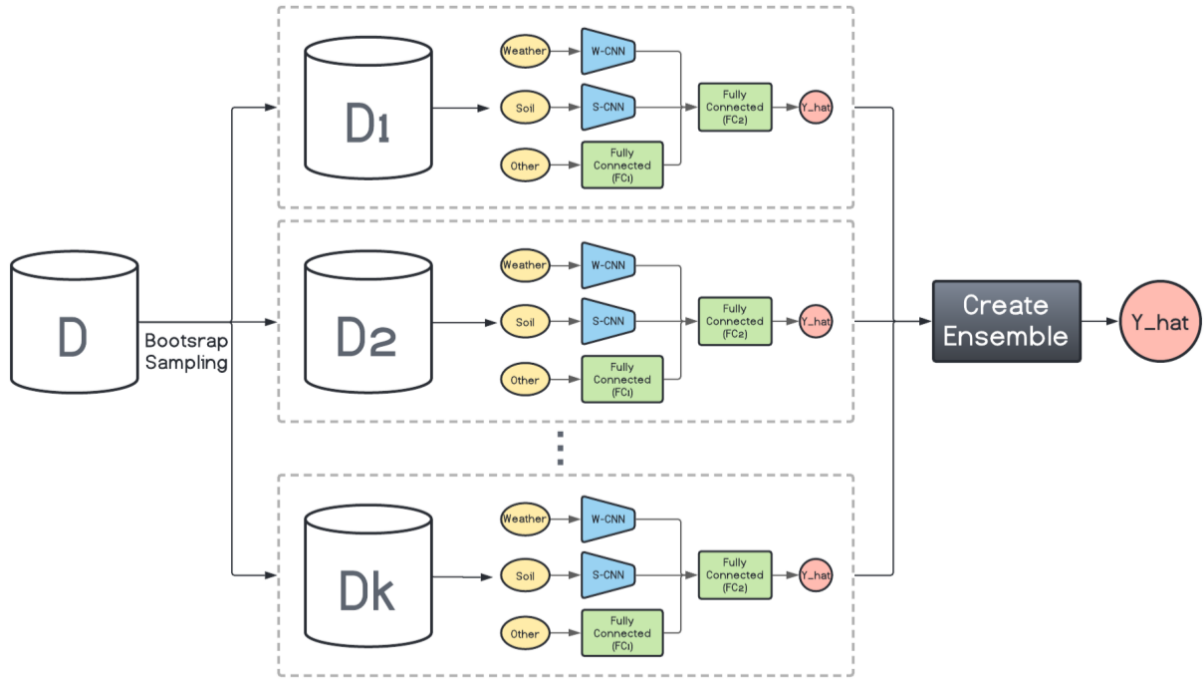

*Figure 2: Homogenous ensemble creation with bagging architecture*
$k$ *data sets (D1, D2, … , Dk) were generated with bootstrap sampling from the original data set (D) and the same base network is trained on each of them. The ensemble creation combines the predictions made by the base networks.*

- *Heterogenous ensembles*

On the other hand, the base models in the heterogenous ensembles are not the same. They can be any machine learning model from the simplest to the most complex models. However, as mentioned before, the ensemble is not expected to perform favorably if the base models do not exhibit a certain level of diversity. To that end, we train $k$ variations of the base CNN-DNN model presented earlier. The foundation architecture of these $k$ models are the same, but the depth level of them are different. In other words, we preserve the same architecture for all models and change the number of features and neurons inside each network to create shallow to deep CNN-DNN models. These models will serve as the inputs to the ensemble creation methods explained in the next chapter.

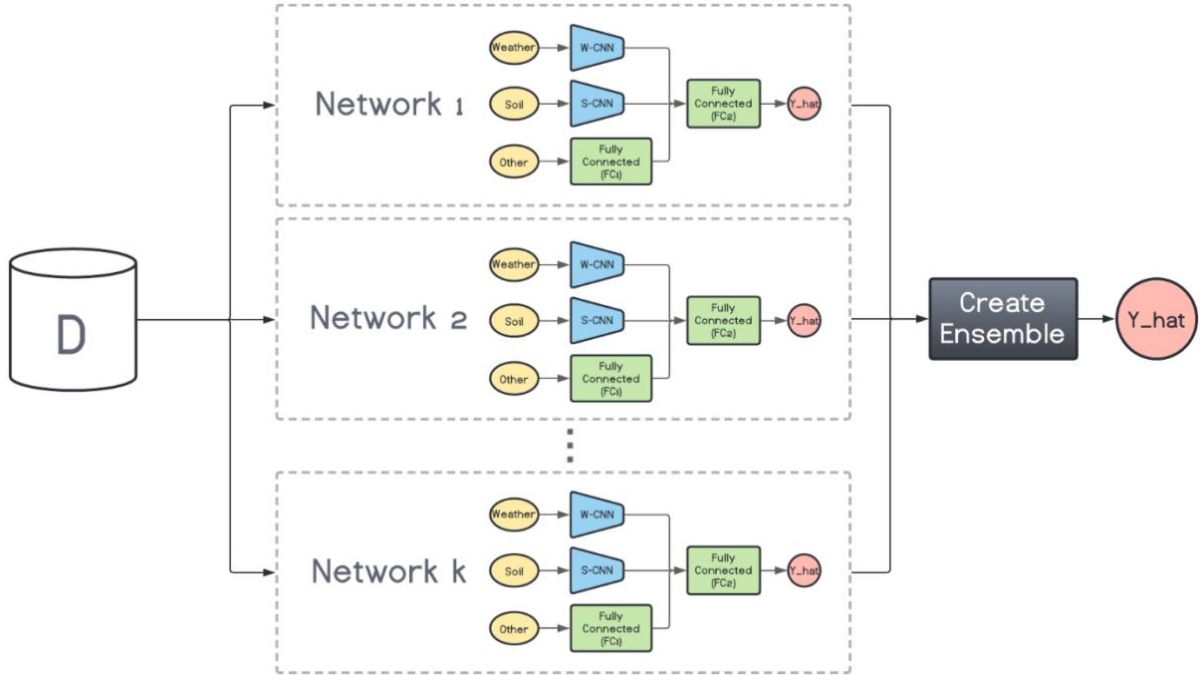

*Figure 3: Heterogenous ensemble creation*
*$k$ networks with the same architecture but with different levels of depth are created using the original data set (D)*

## 2.3. Ensemble Creation

After generating base learners in either of the heterogenous and homogenous methods, they should be combined using a systematic procedure. We have used three different types of ensemble creation methods which are Basic Ensemble Method (BEM), Generalized Ensemble Method (GEM), and stacked generalized ensemble method.

- *Basic Ensemble Method (BEM)*

Perrone and Cooper (1992) proposed BEM as the most natural way of combining base learners. BEM creates a regression ensemble by simple averaging the base estimators. This study claims that BEM can reduce mean squared error of predictions, given that the base learners are diverse.

- *Generalized Ensemble Method (GEM)*

GEM is the general case of a BEM ensemble creation method and tries to create a regression ensemble as the linear combination of the base estimators. Cross-validation is used to generate out-of-bag predictions and optimize the ensemble weights and the model was claimed to avoid overfitting the data (Perrone and Cooper, 1992).

The nonlinear convex optimization problem is as follows.

$$Min \ \tfrac{1}{n}\sum_{i=1}^{n}(y_i - \sum_{j=1}^{k} w_j \hat{y}_{ij})^2 \qquad (2)$$
$$s.t.$$

$$\sum_{j=1}^{k} w_j = 1,$$
$$w_j \geq 0, \quad \forall j = 1, \dots, k.$$

In which $w_j$ is the weight of base model $j$ ($j = 1, \dots, k$), $n$ is the total number of observations, $y_i$ is the true value of observation $i$, and $\hat{y}_{ij}$ is the prediction of observation $i$ by base model $j$.

- *Stacked generalized ensemble method*

Stacked generalization is referred to combining several base estimators by performing at least one more level of machine learning task. Usually, cross-validation is used to generate out-of-bag predictions form the training samples and learn the higher-level machine learning models (Wolpert, 1992). The second level learner can be any choice of ML models. In this study we have selected linear regression, LASSO, random forest and LightGBM as the second level learners.

## 3. Results

The historical county-level data of the US Corn Belt states (Illinois, Indiana, Iowa, Kansas, Michigan, Minnesota, Missouri, Nebraska, North Dakota, Ohio, South Dakota, and Wisconsin) spanning across years 1980-2019 were used to train all considered models. The data from the years 2017, 2018, and 2019, in turn, were reserved as the test data and the data from the years before each of them formed the training data.

As mentioned in the section 2.3, the ensemble creation methods require out-of-bag (OOB) predictions from all the input models that represent the test data to optimally combine the base models. The current procedure to create these OOB predictions is using a cross-validation method. However, due to time-dependency in the training data and the fact that in the homogenous ensemble models the training data is resampled $k$ times, it is not possible to find a consistent vector of OOB predictions across all models and use it to combine the base models. Therefore, 20% of the training data was considered as the validation data and was not used in model training. It is noteworthy that the training data is split to %20-%80 with a stratified split procedure to ensure the validation data has a similar distribution with the training data. To achieve the stratified splits, we binned the observations in the training data into 5 linearly spaced bins based on their corresponding yield values.

The CNN structure of the base models trained for creating homogenous ensemble models are same as the one shown in Table 1. We have resampled the training data 10 times (with replacement) and trained the same CNN-DNN model on each of the 10 newly created training data. The OOB predictions are the predictions made by each of the 10 mentioned models on the validation data.

On the other hand, the base models trained for creating heterogenous ensemble models are not the same and they differ in their CNN depth levels. We trained 5 different CNN-DNN base models on the same training data and formed the OOB predictions by each of those 5 models predicting the observations in the validation data. The details of the CNN components in these 5 models are shown in the Table 2.

*Table 2: Detailed structure of the CNN networks of CNN components designed for heterogenous ensemble models*
*The tables on the left show the details of the CNNs designed for each weather feature, and the right tables present the ones for the CNNs designed for each soil feature. FS, NF, S, and P represent filter size, number of features, stride, and padding.*

| CNNs in the W-CNN component of Model 1 | | | | |
|---|---|---|---|---|
| INPUT SIZE | $32 \times 1$ | | | |
| LAYER NAME | FS | NF | S | P |
| CONV1 | 6 | 2 | 1 | valid |
| AVERAGE POOLING 1 | 2 | - | 2 | valid |
| CONV2 | 3 | 2 | 1 | valid |
| AVERAGE POOLING 2 | 2 | - | 2 | valid |
| CONV3 | 3 | 2 | 1 | valid |
| AVERAGE POOLING 3 | 2 | - | 2 | valid |
| OUTPUT SIZE | $2 \times 1$ | | | |

| CNNs in the S-CNN component of Model 1 | | | | |
|---|---|---|---|---|
| INPUT SIZE | $10 \times 1$ | | | |
| LAYER NAME | FS | NF | S | P |
| CONV1 | 3 | 2 | 1 | valid |
| AVERAGE POOLING 1 | 2 | - | 2 | valid |
| CONV2 | 3 | 2 | 1 | valid |
| AVERAGE POOLING 2 | 2 | - | 2 | valid |
| CONV3 | 3 | 2 | 1 | valid |
| OUTPUT SIZE | $2 \times 1$ | | | |

| CNNs in the W-CNN component of Model 2 | | | | |
|---|---|---|---|---|
| INPUT SIZE | $32 \times 1$ | | | |
| LAYER NAME | FS | NF | S | P |
| CONV1 | 6 | 3 | 1 | valid |
| AVERAGE POOLING 1 | 2 | - | 2 | valid |
| CONV2 | 3 | 3 | 1 | valid |
| AVERAGE POOLING 2 | 2 | - | 2 | valid |
| CONV3 | 3 | 3 | 1 | valid |
| AVERAGE POOLING 3 | 2 | - | 2 | valid |
| OUTPUT SIZE | $3 \times 1$ | | | |

| CNNs in the S-CNN component of Model 2 | | | | |
|---|---|---|---|---|
| INPUT SIZE | $10 \times 1$ | | | |
| LAYER NAME | FS | NF | S | P |
| CONV1 | 3 | 3 | 1 | valid |
| AVERAGE POOLING 1 | 2 | - | 2 | valid |
| CONV2 | 3 | 3 | 1 | valid |
| AVERAGE POOLING 2 | 2 | - | 2 | valid |
| CONV3 | 3 | 3 | 1 | valid |
| OUTPUT SIZE | $3 \times 1$ | | | |

| CNNs in the W-CNN component of Model 3 | | | | |
|---|---|---|---|---|
| INPUT SIZE | $32 \times 1$ | | | |
| LAYER NAME | FS | NF | S | P |
| CONV1 | 6 | 4 | 1 | valid |
| AVERAGE POOLING 1 | 2 | - | 2 | valid |
| CONV2 | 3 | 4 | 1 | valid |
| AVERAGE POOLING 2 | 2 | - | 2 | valid |
| CONV3 | 3 | 4 | 1 | valid |
| AVERAGE POOLING 3 | 2 | - | 2 | valid |
| OUTPUT SIZE | $4 \times 1$ | | | |

| CNNs in the S-CNN component of Model 3 | | | | |
|---|---|---|---|---|
| INPUT SIZE | $10 \times 1$ | | | |
| LAYER NAME | FS | NF | S | P |
| CONV1 | 3 | 4 | 1 | valid |
| AVERAGE POOLING 1 | 2 | - | 2 | valid |
| CONV2 | 3 | 4 | 1 | valid |
| AVERAGE POOLING 2 | 2 | - | 2 | valid |
| CONV3 | 3 | 4 | 1 | valid |
| OUTPUT SIZE | $4 \times 1$ | | | |

| CNNs in the W-CNN component of Model 4 | | | | |
|---|---|---|---|---|
| INPUT SIZE | $32 \times 1$ | | | |
| LAYER NAME | FS | NF | S | P |
| CONV1 | 6 | 5 | 1 | valid |
| AVERAGE POOLING 1 | 2 | - | 2 | valid |
| CONV2 | 3 | 5 | 1 | valid |
| AVERAGE POOLING 2 | 2 | - | 2 | valid |
| CONV3 | 3 | 5 | 1 | valid |
| AVERAGE POOLING 3 | 2 | - | 2 | valid |
| OUTPUT SIZE | $5 \times 1$ | | | |

| CNNs in the S-CNN component of Model 4 | | | | |
|---|---|---|---|---|
| INPUT SIZE | $10 \times 1$ | | | |
| LAYER NAME | FS | NF | S | P |
| CONV1 | 3 | 5 | 1 | valid |
| AVERAGE POOLING 1 | 2 | - | 2 | valid |
| CONV2 | 3 | 5 | 1 | valid |
| AVERAGE POOLING 2 | 2 | - | 2 | valid |
| CONV3 | 3 | 5 | 1 | valid |
| OUTPUT SIZE | $5 \times 1$ | | | |

| CNNs in the W-CNN component of Model 5 | | | | |
|---|---|---|---|---|
| INPUT SIZE | $32 \times 1$ | | | |
| LAYER NAME | FS | NF | S | P |
| CONV1 | 6 | 6 | 1 | valid |
| AVERAGE POOLING 1 | 2 | - | 2 | valid |
| CONV2 | 3 | 6 | 1 | valid |
| AVERAGE POOLING 2 | 2 | - | 2 | valid |
| CONV3 | 3 | 6 | 1 | valid |
| AVERAGE POOLING 3 | 2 | - | 2 | valid |
| OUTPUT SIZE | $6 \times 1$ | | | |

| CNNs in the S-CNN component of Model 5 | | | | |
|---|---|---|---|---|
| INPUT SIZE | $10 \times 1$ | | | |
| LAYER NAME | FS | NF | S | P |
| CONV1 | 3 | 6 | 1 | valid |
| AVERAGE POOLING 1 | 2 | - | 2 | valid |
| CONV2 | 3 | 6 | 1 | valid |
| AVERAGE POOLING 2 | 2 | - | 2 | valid |
| CONV3 | 3 | 6 | 1 | valid |
| OUTPUT SIZE | $6 \times 1$ | | | |

To evaluate the performance of the trained heterogenous and homogenous CNN-DNN ensembles, the ensembles created from five individual machine learning models (linear regression, LASSO, XGBoost, random forest, and LightGBM) were considered as benchmark and were trained on the same data sets developed for training the CNN-DNN ensemble models. The benchmark models were run on a computer equipped with a 2.6 GHz Intel E5-2640 v3 CPU, and 128 GB of RAM. The CNN-DNN models were run on a computer with a 2.3 GHz Intel E5-2650 v3 CPU, NVIDIA k20c GPU, and 768 GB of RAM.

The predictive performance of these ensemble models was previously shown in two separate published papers (Shahhosseini et al., 2020; Shahhosseini et al., 2021). The results are summarized in the Table 3 (See Figure S3 for XY plots of some of the designed ensembles).

Table 3: Test prediction error (RMSE) and coefficient of determination ($R^2$) of designed ensemble models compared to the benchmark ensembles (Shahhosseini et al., 2020; Shahhosseini et al., 2021).

| ML models | BEM | | GEM | | Stacked regression | | Stacked LASSO | | Stacked random forest | | Stacked LightGBM | |
|---|---|---|---|---|---|---|---|---|---|---|---|---|
| | RMSE (kg/ha) | $R^2$ (%) | RMSE (kg/ha) | $R^2$ (%) | RMSE (kg/ha) | $R^2$ (%) | RMSE (kg/ha) | $R^2$ (%) | RMSE (kg/ha) | $R^2$ (%) | RMSE (kg/ha) | $R^2$ (%) |
| Test year: 2017 - Training years: 1980-2016 | | | | | | | | | | | | |
| Benchmark | 960 | 79.6% | 1002 | 77.7% | 1014 | 77.2% | 1012 | 77.3% | 1024 | 76.7% | 999 | 77.9% |
| Heterogenous | 1003 | 77.7% | 969 | 79.2% | 908 | 81.8% | 908 | 81.7% | 978 | 78.8% | 933 | 80.7% |
| Homogenous | 954 | 79.8% | 944 | 80.3% | 875 | 83.0% | 874 | 83.1% | 936 | 80.6% | 906 | 81.8% |
| Test year: 2018 - Training years: 1980-2017 | | | | | | | | | | | | |
| Benchmark | 1145 | 74.7% | 1047 | 78.8% | 1041 | 79.0% | 1041 | 79.0% | 1101 | 76.6% | 1070 | 77.9% |
| Heterogenous | 1065 | 78.0% | 1094 | 76.8% | 1072 | 77.8% | 1072 | 77.8% | 1116 | 75.9% | 1087 | 77.2% |
| Homogenous | 1033 | 79.4% | 992 | 81.0% | 1058 | 78.4% | 1056 | 78.4% | 1077 | 77.6% | 1065 | 78.1% |
| Test year: 2019 - Training years: 1980-2018 | | | | | | | | | | | | |
| Benchmark | 936 | 72.6% | 1035 | 66.4% | 1028 | 66.9% | 1035 | 66.5% | 1084 | 63.2% | 1029 | 66.9% |
| Heterogenous | 900 | 74.6% | 1083 | 63.3% | 1282 | 48.5% | 1279 | 48.8% | 1225 | 53.0% | 1234 | 52.3% |
| Homogenous | 866 | 76.5% | 867 | 76.5% | 885 | 75.5% | 883 | 75.6% | 932 | 72.8% | 895 | 74.9% |

The heterogenous and homogenous ensemble models both provide improvements over the well-performing ensemble benchmarks in most cases (Table 3). However, the heterogenous ensemble model is constantly outperformed by the homogeneous ensemble models. This is in line with what we expected as the homogeneous model inherently introduces more diversity in the ensemble base models which in turn will result in lowering the prediction variance and consequently better generalizability of the trained model. The performance comparison of homogeneous ensemble model compared to the benchmark is shown in the Figure 4. Another observation in the Table 3 is that in case of homogenous ensembles, some of the ensemble creation methods have made better predictions than average homogeneous ensemble (BEM) i.e., bagged CNN-DNN. This again confirms our assertion that assigning unequal weights to the bagged models results in better predictions.

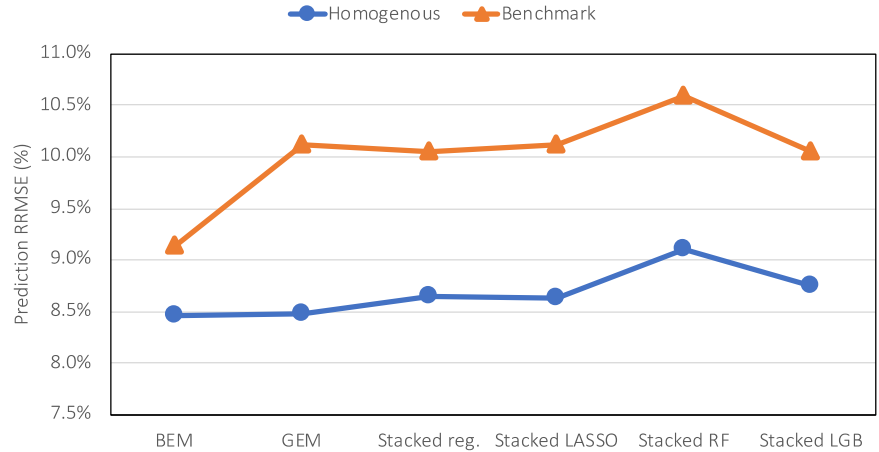

*Figure 4: Comparing prediction error (relative RMSE) of the homogeneous model with the benchmark on the data from the year 2019 taken as the test data*

The generalizability of all trained models is proved as we have shown that in three test scenarios, the ensemble models demonstrate superb prediction performance. This also can be observed by looking at the train and test loss vs. epochs graphs. Some examples of these graphs are shown in Figure 5. As the figure suggests, the dropout layers could successfully prevent overfitting of the CNN-DNN models, and the test errors tend to stay stable across the iterations. The generalizability of the trained models will further be discussed in the chapter 4.

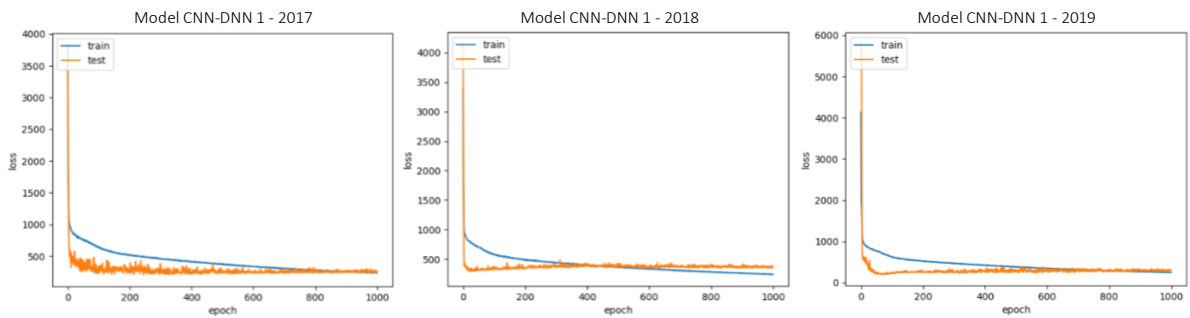

*Figure 5: Train and test loss vs. epochs of some of the trained CNN-DNN models*

## 4. Discussion

### 4.1. Models' performance comparison with the literature

We designed a novel CNN-DNN ensemble model with the objective of providing the most accurate prediction model for county-level corn yield across US Corn Belt states. The numerical results confirmed the superb performance of the designed ensemble models compared to literature models. Table 3 showed that the homogenous ensemble models outperform the benchmark (Shahhosseini et al., 2020) by 10-16%. In addition, comparing the results with another well-performing prediction model in the literature (Khaki et al., 2020b), the homogeneous ensemble could outperform the prediction results of Khaki et al. (2020b) by 10-12% in common test set scenarios (2017 and 2018 test years). The CNN-RNN model developed by Khaki et al. (2020b) presented test prediction errors of 988 kg/ha (15.74 bu/acre) and 1107 kg/ha (17.64 bu/acre) for the test years 2017 and 2018, respectively, while the homogeneous ensemble model designed here resulted in test prediction errors of 874 kg/ha (13.93 bu/acre) and 992 kg/ha (15.8 bu/acre) for the test years 2017 and 2018, respectively.

This is the first study that designed a novel ensemble neural network architecture that has the potential to make the most accurate yield predictions. The model developed here is advantageous compared to the literature due to the ability of the ensemble model in decreasing prediction variance by combining diverse models as well as reducing prediction bias by training the ensemble model based on powerful base models. Shahhosseini et al. (2020) had used ensemble learning for predicting county-level yield prediction, but neural network-based architectures were not considered, and the models were trained only on three states (IL, IA, IN). Khaki et al. (2020b) trained a CNN-RNN model for predicting US Corn Belt corn and soybean yields, but the model developed there is unable to make predictions as accurate as the models designed in this study and is not benefitting from the diversity in the predictions.

Including remote sensing data as well as simulated data from crop model like APSIM could potentially improve the predictions made by our models further which can be pursued as the future research direction. In addition, we assumed all considered farms are rainfed, while in states such as Kansas and Nebraska many of the farms are irrigated. Surprisingly, the prediction accuracy in these states was comparable with other states (Figures 6 and 7). We believe this is because of the use of average or rainfed corn yields from these states, not irrigated yields to train our models. Including the irrigation data can result in better prediction and perhaps new models for those states and is another possible future research direction.

### 4.2. Comparing the models' performance across US Corn Belt states

Figure 6 compares the prediction errors of the test year of 2019 for some of the designed ensemble models represented by relative root mean squared error (RRMSE) for each of the 12 US Corn Belt states under study. The models performed the best in Iowa, Illinois, and Nebraska, and worst in Kansas and South Dakota. The worse prediction error in Kansas can be explained by the fact that the majority of the farms in Kansas state are irrigated and this irrigation is not considered as one of the variables when training the ensemble models. It is clear that including irrigation variable can improve the predictions. However, that was not the case for Nebraska, suggesting that irrigation may not be the only reason for the low performance in Kansas. Upon further investigate, we realized the corn yields in the Nebraska state are highly correlated with the weather features especially maximum temperature, while the corn yields in the Kansas state don't show this amount of correlation to weather features and are slightly correlated with both weather and soil features. In other words, it seems that although the weather features are adequate for making decent predictions in the Nebraska state, this is not the case for the Kansas.

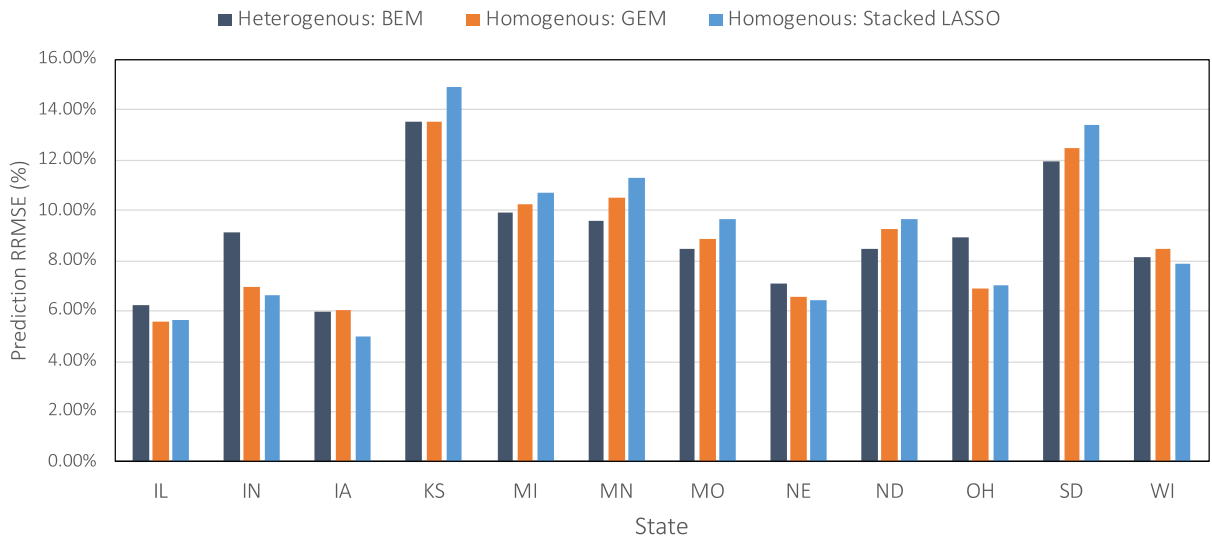

*Figure 6: Comparing prediction error (relative RMSE) of the some of the designed ensembles across all US Corn Belt states on the data from the year 2019 taken as the test data*

Figure 7 depicts the relative error percentage of each year's test predictions on a county choropleth map of the US Corn Belt. The errors are calculated by dividing over/under prediction of the homogenous GEM model divided by the yearly average yield. This figure proves that the model is robust and can be easily generalized to other environments/years. One observation is that the model keeps overpredicting the yields in the Kansas state. This could be explained by the irrigation assumption we made when developing the data set. We assumed all the farms are rainfed and did not consider irrigation in states like Kansas in which some of the farms are irrigated.

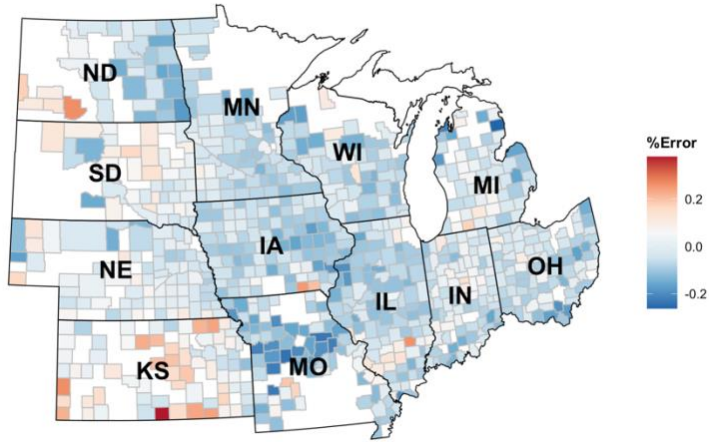
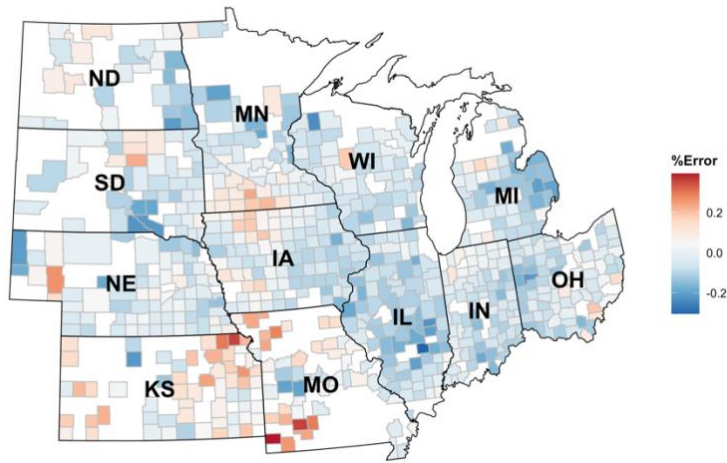
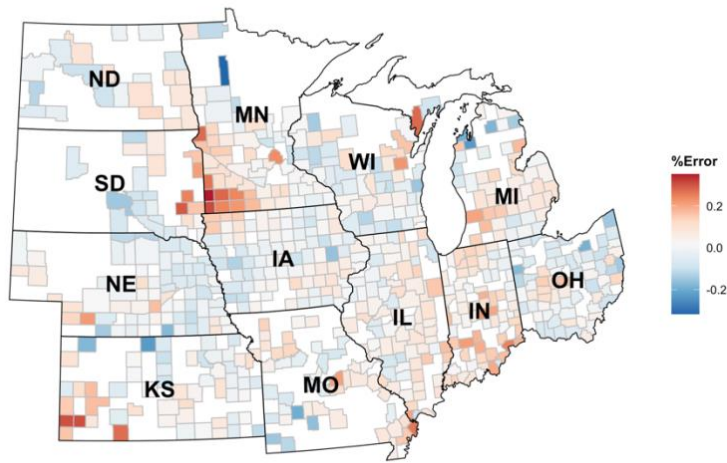

*Figure 7: relative percentage error of the Homogenous GEM predictions shown on a choropleth map of the US Corn Belt*

## 4.3. Generalization power of the designed Ensemble CNN-DNN models

To further test the generalization power of the designed ensembles, we gathered the data of all considered US Corn Belt states for the year 2020 and applied the trained heterogeneous and homogeneous ensemble models as well as the benchmarks on the new unseen observations of the year 2020. As the results imply (Table 4), both heterogenous and homogeneous ensemble models provide better predictions than the benchmark ensemble models, with the homogeneous Generalized Ensemble Model (GEM) being the most accurate prediction model. This model could provide predictions with 958 kg/ha root mean squared error and explain about 77% of the total variability in the response variable.

Table 4: Test prediction error (RMSE) and coefficient of determination ($R^2$) of designed ensemble models compared to the benchmark ensembles (Shahhosseini et al., 2020; Shahhosseini et al., 2021) when applied on 2020 test data

| ML models | BEM | | GEM | | Stacked regression | | Stacked LASSO | | Stacked random forest | | Stacked LightGBM | |
|---|---|---|---|---|---|---|---|---|---|---|---|---|
| | RMSE (kg/ha) | $R^2$ (%) | RMSE (kg/ha) | $R^2$ (%) | RMSE (kg/ha) | $R^2$ (%) | RMSE (kg/ha) | $R^2$ (%) | RMSE (kg/ha) | $R^2$ (%) | RMSE (kg/ha) | $R^2$ (%) |
| Test year: 2020 - Training years: 1980-2018 | | | | | | | | | | | | |
| Benchmark | 1115 | 68.4% | 1165 | 65.5% | 1166 | 65.4% | 1170 | 65.2% | 1210 | 62.8% | 1183 | 64.4% |
| Heterogenous | 972 | 76.0% | 989 | 75.1% | 992 | 75.0% | 991 | 75.0% | 1048 | 72.1% | 1000 | 74.6% |
| Homogenous | 982 | 75.5% | 958 | 76.7% | 1001 | 74.5% | 999 | 74.6% | 1053 | 71.8% | 1018 | 73.6% |

## 5. Conclusion

In this study we designed two novel CNN-DNN ensemble types for predicting county-level corn yields across US Corn Belt states. The base architecture used for creating the ensembles is a combination of convolutional neural networks and deep neural networks. The CNNs were in charge of extracting useful high-level features from the soil and weather data and provide them to a fully connected network for making the final yield predictions. The two ensemble types were heterogeneous and homogeneous which used the same base CNN-DNN structure but generated the base models in different manners. The homogenous ensemble used one fixed CNN-DNN network but applied it on multiple bagged data sets. The bagged data sets introduced a certain level of diversity that the created ensembles had benefited from. On the other hand, the heterogeneous ensemble used different base CNN-DNN networks which shared the same structure but differed in their depth levels. The different depth levels were considered as another method of introducing diversity into the ensembles. All base models generated from either of these two ensemble types were combined with each other using three ensemble creation methods: Basic Ensemble Method (BEM), Generalized Ensemble Method (GEM), and stacked generalized ensembles. The numerical results showed that the ensemble models of both homogeneous and heterogeneous types could outperform the benchmark ensembles which had previously proved to be effective (Shahhosseini et al., 2020, Shahhosseini et al., 2021) as well as well-performing CNN-RNN architecture designed by Khaki et al. (2020b). In addition, homogeneous ensembles provide the most accurate predictions across all US Corn Belt states. The results demonstrated that in addition to the fact that these ensemble models benefitted from higher level of diversity from the bagged data sets, they provided a better combination of base models compared to simple averaging in the bagging. The generalization power of the designed ensembles was proved by applying them on the unseen observations of the year 2020. Once again heterogeneous and homogeneous ensemble models outperformed the benchmark ensembles.